# Quantitative investigation of the inverse Rashba-Edelstein effect in Bi/Ag and Ag/Bi on YIG

Masasyuki Matsushima, Yuichiro Ando, Sergey Dushenko, Ryo Ohshima,
Ryohei Kumamoto, Teruya Shinjo and Masashi Shiraishi [#]

*Department of Electronic Science and Engineering,*
*Kyoto University, Kyoto 615-8510, Japan.*

[#] Corresponding author: Masashi Shiraishi (mshiraishi@kuee.kyoto-u.ac.jp)

**Abstract**

The inverse Rashba-Edelstein effect (IREE) is a spin conversion mechanism that recently attracts attention in spintronics and condensed matter physics. In this letter, we report an investigation of the IREE in Bi/Ag by using ferrimagnetic insulator yttrium iron garnet (YIG). We prepared two types of samples with opposite directions of the Rashba field by changing a stacking order of Bi and Ag. An electric current generated by the IREE was observed from both stacks, and an efficiency of spin conversion—characterized by the IREE length—was estimated by taking into account a number of contributions left out in previous studies. This study provides a further insight into the IREE spin conversion mechanism: important step towards achieving efficient spin-charge conversion devices.

The conversion from a spin current to a charge current and vice versa, has been attracting growing interest over the last few years. One of the main mechanisms for conversion is the intrinsic spin-orbit interaction of the material, which is roughly proportional to $Z^4$ ($Z$ is the atomic number). Thus, heavy elements, such as Pt [1], Ta [2], W [3] and Bi [4-6], have been intensively studied, resulting in a detection of the inverse spin Hall effect (ISHE) in heavy metals [7]. In the ISHE, the spin conversion efficiency is described as the spin Hall angle. The large spin Hall angle is generally observed for heavy metals in the agreement with $Z^4$ trend [8]. However, one cannot continue to increase the spin Hall angle by moving to heavier elements, since materials with large atomic number tend to be unstable. In this context, a quest arose for a new spin conversion mechanism and material with high spin conversion efficiency.

In 2013, attention was drawn to a new spin conversion mechanism: the inverse Rashba-Edelstein effect (IREE), which was firstly reported in a 2-dimenstional electron gas [9, 10] and was later rediscovered at a hetero-interface of Bi/Ag [11], where a large Rashba splitting gives rise to the spin-to-charge conversion [12]. The Rashba field causes spin splitting of the dispersion curve of free electrons: at the Fermi surface, electrons are restricted to the in-plane direction and move according to their spin direction, like spin-momentum-locking in topological insulators. Since then, the IREE has been measured in the other systems, such as $\alpha$-Sn [13] and LaAlO/SrTiO [14]. The spin conversion efficiency of the IREE can exceed that of the known ISHE materials [13], which may allow one to fabricate efficient spin convertor devices.

Up to now, in all of the IREE experiments, conductive ferromagnets such as NiFe [11] and Fe [15] have been used as spin injectors. Especially, the stacking order of Bi and Ag was changed to reverse the direction of the Rashba field and the sign reversal of the IREE length, the efficiency index of the IREE, was observed [15], which was a missing experiment in ref. 11 and provided another evidence of the IREE in the Bi/Ag system. In such experiments, an electric current due to the IREE generates electromotive force that is experimentally detected under ferromagnetic resonance (FMR) of the ferromagnet. However, recent studies revealed that conductive ferromagnets generate spurious electromotive forces under the FMR due to the planar Hall effect [16], the self-induced inverse spin Hall effect (NiFe [17] and Fe [18]) and magnonic charge pumping [19]. The electric current generated by these unwanted effects is detected together with the current generated by the IREE, which hinders precise estimation of the spin conversion efficiency of the IREE, i.e., the IREE length. To avoid these spurious effects superposing the IREE, an introduction of insulating spin source is indispensable. Furthermore, spin current dissipation in both Bi and Ag due to the ISHE and transparency of spin current at the Bi/Ag interface were not fully considered in previous IREE studies, which can also impede precise estimation of the IREE length.

In this letter, we introduce ferrimagnetic insulator yttrium iron garnet (YIG) as a spin injector instead of conductive ferromagnet, which eliminates spurious electromotive forces from the ferromagnet layer. In addition, two types of samples with different stacking orders—Bi/Ag/YIG and Ag/Bi/YIG—were studied to understand an effect of the sign reversal of the Rashba field. To estimate the IREE length, we also took into account spin current dissipation in Bi and Ag layers due to the ISHE.

We prepared two types of samples with different directions of the Rashba fields: Bi/Ag/YIG and Ag/Bi/YIG. Bi (7-nm-thick) and Ag (5-nm-thick) were deposited on single crystal YIG (10-μm-thick layer) by a resistance heating method (see Fig. 1(a)). Whereas the direction of the Rashba field with respect to the spin direction determines the sign of the IREE current, only Py/Ag/Bi/SiO$_2$ stacking order was investigated in the previous study of the IREE at the interface between Bi and Ag [11]. The samples with opposite directions of the Rashba field in this study allowed further understanding of its role in the IREE. We used spin pumping to inject a pure spin current from the YIG into the Bi/Ag and the Ag/Bi. Spin pumping is a method of spin injection from a ferromagnet (or ferrimagnet) to an adjacent layer under the FMR of the ferromagnet. Under the FMR, a part of the spin angular momentum in the ferromagnet is transferred to the carriers in the adjacent layer via *s-d* coupling, resulting in spin accumulation and generation of a pure spin current in the latter [7]. In the spin pumping measurements, the Bi/Ag/YIG and the Ag/Bi/YIG samples were placed in a TE$_{011}$ (transverse electric mode) cavity of an electron spin resonance system (JEOL JES-FA 200), the microwave frequency and power were set to be 9.12 GHz and 1 mW, respectively (Fig. 1(b)). All measurements were carried out at room temperature.

We clarified crystal structures of the Bi and the Ag using an X-ray diffraction *θ-2θ* patterns of the Bi (20 nm)/Ag (30 nm) and the Ag (30 nm)/Bi (20 nm), where Cu-Kα radiation was used. Figures 1(c) and 1(d) show the X-ray diffraction patterns of the Bi/Ag/YIG and the Ag/Bi/YIG. The similar diffraction pattern from NiFe/Ag/Bi/SiO$_2$—the peaks Bi(003) and Ag(220)—was measured in ref. [11]. This suggests that the crystallography of the Ag/Bi/YIG and the NiFe/Ag/Bi/SiO$_2$ is similar, and the existing Rashba fields in the current and the previous studies expected to be of similar magnitude. On the contrary, X-ray diffraction pattern of the Bi/Ag/YIG stack showed the peaks Ag(111) and Bi(012).

Figures 2(a) and 2(b) show electric currents generated from the Bi/Ag/YIG and the Ag/Bi/YIG samples under the FMR of the YIG. A spin conversion signal was obtained after subtracting data for the opposite directions of the external magnetic field to eliminate any heat-related spurious effects. The IREE is an interface-induced effect, whereas the ISHE is a bulk-induced effect. Thus, we show a magnitude of an electric current instead of an electromotive force, because the electromotive force is affected by a conductivity and thickness of samples. The electric current was generated at the FMR point in both samples: the magnitude was 300 and 200 pA for the Bi/Ag/YIG and the Ag/Bi/YIG, respectively. Taking into account the positive spin Hall angle of Bi [4] and Ag [8], and the direction of the Rashba field at the Bi/Ag interface [11], the observed negative electric current indicates in the spin conversion via electron carriers in our measurement setup. The observed difference in the external magnetic field dependence of the generated electric current around the FMR point cannot be ascribed to the slight difference in the FMR spectra of the YIG (see supplementary material). Below we discuss two key features of the data. First of all, we observed enhanced electric current in the Ag/Bi/YIG comparing with that in Bi/YIG. A previous study on spin conversion in Bi on YIG reported that electric current generated in Bi due to the ISHE was less than 170 pA for the 10-nm-thick Bi layer (when the Bi thickness was less than 10 nm, the intensity of the electric current was below the detection level) [6]. Since Bi thickness in this study is 7 nm, the measured electric current from the Ag/Bi/YIG cannot be attributed only to the ISHE of the Bi layer. Second key feature is the same polarity of the electric current generated under spin pumping for both types of the samples: Ag/Bi/YIG and Bi/Ag/YIG. Polarity of the electric current in the Ag/Bi/YIG was not reversed, although the direction of the Rashba field is opposite for the two types of samples. To interpret the experimental data, we constructed a model that includes generation of the spin-charge conversion current by both the IREE and the ISHE, which was partly considered in the study [15], and also considers spin transparency, $\kappa$, of the Ag/Bi interface. The IREE lengths calculated for the $\kappa = 0$ and $\kappa = 1$ are the upper and the lower limits, respectively. For $\kappa = 0$ (see Fig. 3(a)), the spin current (injected from the YIG) dissipates due to the ISHE, while it diffuses through the Bi (or the Ag) layer. However, the Bi/Ag is considered completely non-transparent for spin current ($\kappa = 0$), i.e. 100% of the spin current is dissipated at the Bi/Ag interface. For $\kappa = 1$, we assume that the spin current does not dissipate at the Bi/Ag interface, it is constantly supplied from the YIG layer (note that spin accumulation, not spin-flip scattering, gives rise to the IREE), and it is reflected at the surface of the sample (Fig. 3(b)). Thus, we also take into account the ISHE-induced spin-charge conversion in the Bi and the Ag layers due to the reflected spin current. The generated under spin pumping electric current for a given $\kappa$ value is given by

$$l\theta_{\text{SHE}}(a)\lambda_a \left(1 - e^{-\frac{t_a}{\lambda_a}}\right)\left(1 - \kappa^2 e^{-\frac{t_a}{\lambda_a} - \frac{2t_b}{\lambda_b}}\right) j_s^0 \left(\frac{2e}{\hbar}\right) \pm l\lambda_{\text{IREE}} e^{-\frac{t_a}{\lambda_a}} \left(1 + \kappa e^{-\frac{2t_b}{\lambda_b}}\right) j_s^0 \left(\frac{2e}{\hbar}\right) +$$

$$\kappa l\theta_{\text{SHE}}(b)\lambda_b e^{-\frac{t_a}{\lambda_a}} \left(1 - e^{-\frac{t_b}{\lambda_b}}\right)^2 j_s^0 \left(\frac{2e}{\hbar}\right) = I_c,$$

where $a$, $b$ = Bi or Ag, with $a$ being a layer evaporated directly on top of the YIG substrate, $l$ is the sample length (2 mm), $j_s^0$ is the spin current density at the YIG/$a$ interface, $e$ is elementary charge, $\hbar$ is Dirac constant, $t_{\text{Ag}}$ is the thickness of Ag (5 nm), $\lambda_{\text{Ag}}$ is the spin diffusion length in Ag (200 nm [20]), $\theta_{\text{SHE}}(\text{Ag})$ is the spin Hall angle of Ag (0.007 [21]), $t_{\text{Bi}}$ is the thickness of Bi (7 nm), $\lambda_{\text{Bi}}$ is the spin diffusion length in Bi (8 nm [5]), $\theta_{\text{SHE}}(\text{Bi})$ is the spin Hall angle of Bi (0.02 [5]), and $\lambda_{\text{IREE}}$ is the inverse Rashba-Edelstein length. The sign before the second term corresponds to the direction of the Rashba field: "+" for the Bi/Ag/YIG stack, and "-" for the Ag/Bi/YIG stack. Using the experimentally detected spin-charge conversion current $I_c$ of 300 and 200 pA for direct and reversed stacks, and the above equation, the IREE lengths for $\kappa = 0$ and 1 were estimated to be 0.07 and 0.001 nm, respectively. Figure 4 shows the IREE length dependence on the $\kappa$ in the [0,1] range. The IREE length estimated in our study is at least 30 times smaller than that in the previous studies (~0.3 nm [11,15]), and is slightly larger or comparable to those in the other studies using Bi/Cu (0.009 nm [22]) and Sb/Ag (0.01 nm [23]). However, simultaneously, it is noted that NiFe or Fe was used as a spin source of spin pumping in ref. [11,15,23]. We note that IREE length can depend on the interface quality, and, thus, vary between different studies. However, in contrast to previous studies, we excluded contribution to the electromotive force from the voltage induced in the conductive ferromagnets, and also took into account the electromotive force generated due to the ISHE. In the study of the annihilation rate of the positronium states in the surface layers of the Bi/Ag and Ag/Bi, experimental data were interpreted as a signature of the opposite spin polarization generated by Rashba-Edelstein effect (the effect reciprocal the IREE) for different stacking order of the layers [25]. Since all three studies ([11,24] and present work) reported similar crystal structure of the Bi and Ag layer from the X-ray diffraction measurements (Bi(003), Bi(012) and Ag(220) peaks), how sample quality affects ratio between IREE and ISHE remains an open question. To provide a material for discussing a contribution of sample quality, we also carried out the similar experiment using Bi/Ag and Ag/Bi on NiFe, which is described in supplementary material.

Finally, we discuss limitations of the used model. The amplitude of the Rashba field (but not the sign), the spin-transparency of the Bi/Ag interface and the spin Hall angles of the layers were assumed to be the same for Bi/Ag/YIG and Ag/Bi/YIG stacks. As can be seen from the X-ray diffraction data, crystal structures, and thus, probably quality of the interface are not identical for the two types of the stacks. However, we stress that the main experimental result—the same polarity of the generated voltage for the Bi/Ag/YIG and Ag/Bi/YIG—holds true even when quality of the interfaces differs. While interface quality can influence amplitude of the Rashba field, its direction should be reversed together with the stacking order of the Bi and Ag layers. Thus, our study shows that ISHE-induced spin-charge conversion is non-negligible in the Bi/Ag system. As for the direction for the future studies of the IREE length, it is necessary to (1) measure spin-transparency of the Bi/Ag interface, and (2) achieve the same quality of Bi and Ag layers for any growth order of the layers. However, since underlayer of the Bi/Ag interface is always different for the reversed stacking order, fully identical interfaces cannot be obtained experimentally.

In summary, we investigated the IREE in Bi/Ag using spin pumping from the ferrimagnetic insulator YIG. We observed the same polarity of the spin-charge conversion current for Bi/Ag/YIG and Ag/Bi/YIG stacks, in contrast to the expectation from the simple phenomenological picture. The experimental results were explained by the model that takes into account spin-charge conversion due to both the IREE and the ISHE effects. The IREE length of Bi/Ag in our samples was estimated to be in the range 0.001-0.07 nm.

**Supplementary Material**

See supplementary material for FMR spectra from the YIG of Bi/Ag/YIG and Ag/Bi/YIG samples and results of spin conversion measurements of Bi/Ag/Py and Ag/Bi/Py on quartz substrates.

**Acknowledgements**

This research was supported in part by a Grant-in-Aid for Scientific Research from the Ministry of Education, Culture, Sports, Science and Technology (MEXT) of Japan, Innovative Area "Nano Spin Conversion Science" (No. 26103003), Scientific Research (S) "Semiconductor Spincurrentronics" (No. 16H0633) and Scientific Research (A) "Multiferroics in Dirac electron materials" (No. 15H02108), and also by Kyoto University Nano Technology Hub in "Nanotechnology Platform Project" sponsored by the Ministry of Education, Culture, Sports, Science and Technology (MEXT), Japan. S.D. acknowledges support by JSPS Postdoctoral Fellowship and JSPS KAKENHI Grant No. 16F16064.

**Figure Captions**

**Fig. 1. (a)** Structure of the samples. Direction of the Rashba field is expected to be opposite for the reversed stacking order. **(b)** Measurement setup of the IREE at the Bi/Ag interface using spin pumping. **(c)** and **(d)** X-ray spectra of Bi (20 nm)/Ag (30 nm)/YIG and Ag (30 nm)/Bi (20 nm)/YIG, respectively.

**Fig. 2. (a)** An electric current observed from the Bi/Ag/YIG sample. The magnitude of the current was calculated to be 300 pA. **(b)** An electric current observed from the Ag/Bi/YIG sample. The magnitude of the current was calculated to be 200 pA. Generated electric current for both stacks had maximum at the FMR magnetic field, consistent with the spin pumping spin injection. Polarity of the electric current was the same for both stacks, i.e. for opposite Rashba field directions.

**Fig. 3.** Schematic illustrations of the used spin-charge conversion model for $\kappa = 0$ and $\kappa = 1$. In both cases, the ISHE-induced spin relaxation and spin conversion in Bi and Ag are taken into account to estimate the IREE length. **(a)** $\kappa = 0$: 100% of the spin current from the YIG is dissipated at the Bi/Ag interfaces. **(b)** $\kappa = 1$: Bi/Ag interface is fully spin-transparent, in addition, the spin current is reflected at the surface of the sample.

**Fig. 4.** IREE length $\lambda_{IREE}$ as a function of spin transparency $\kappa$ in the [0,1] range. An increase in $\kappa$ leads to an increase in the contribution from the ISHE from the layer that is not contact with YIG, as a result, the value of the IREE length decreases. The maximum ($\kappa = 0$) and minimum ($\kappa = 1$) of $\lambda_{IREE}$ are 0.07 nm and 0.001 nm, respectively.

**Figures**

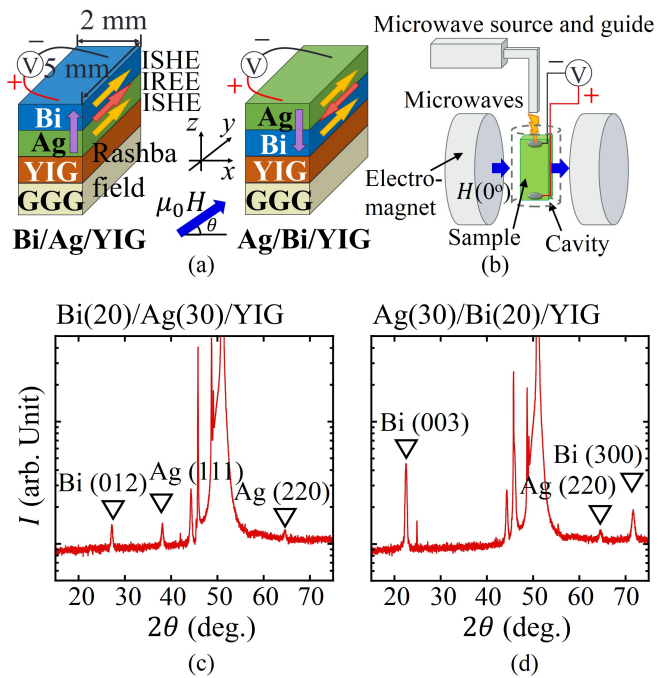

Fig. 1 M. Matsushima et al.

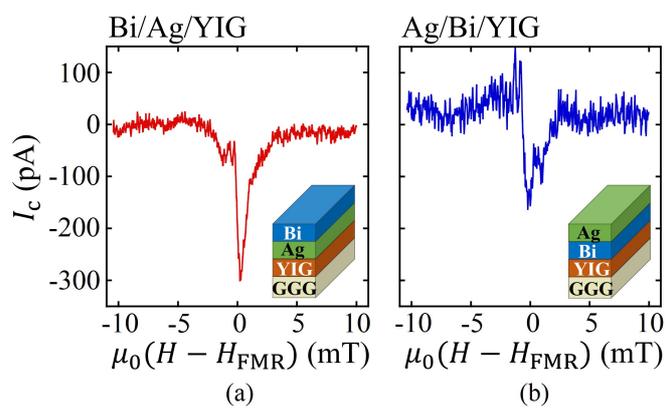

Fig. 2 M. Matsushima et al.
12

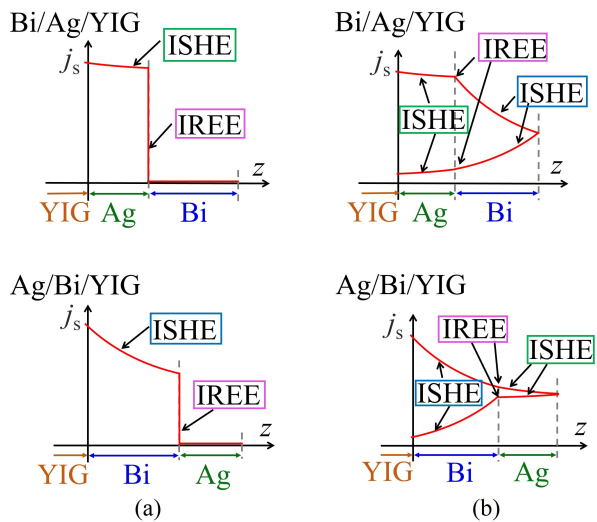

Fig. 3 M. Matsushima et al.

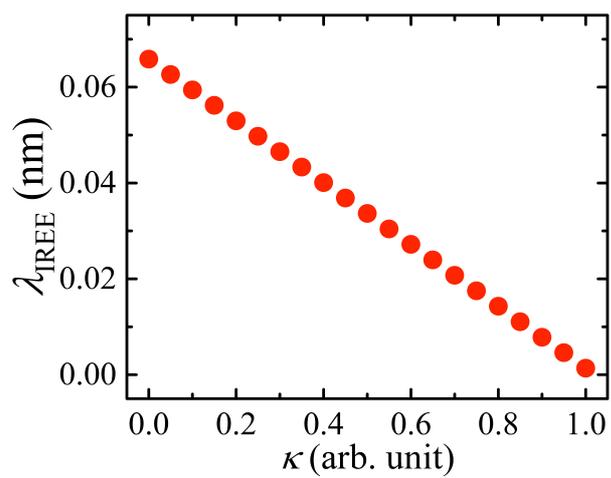

Fig. 4 M. Matsushima et al.

# Quantitative investigation of the inverse Rashba-Edelstein effect in Bi/Ag and Ag/Bi on YIG

Masasyuki Matsushima, Yuichiro Ando, Sergey Dushenko, Ryo Ohshima, Ryohei Kumamoto, Teruya Shinjo and Masashi Shiraishi [#]

*Department of Electronic Science and Engineering,*
*Kyoto University, Kyoto 615-8510, Japan.*

### A. FMR spectra of YIG

Fig. S1 shows the FMR spectra for Bi/Ag/YIG and Ag/Bi/YIG. The in-plane external magnetic field was applied in the 0 and 180 degrees directions. As expected from the YIG FMR signal, the shape and amplitude of the FMR signals were closely the same for both types of the stacks, and for both directions of the external magnetic field.

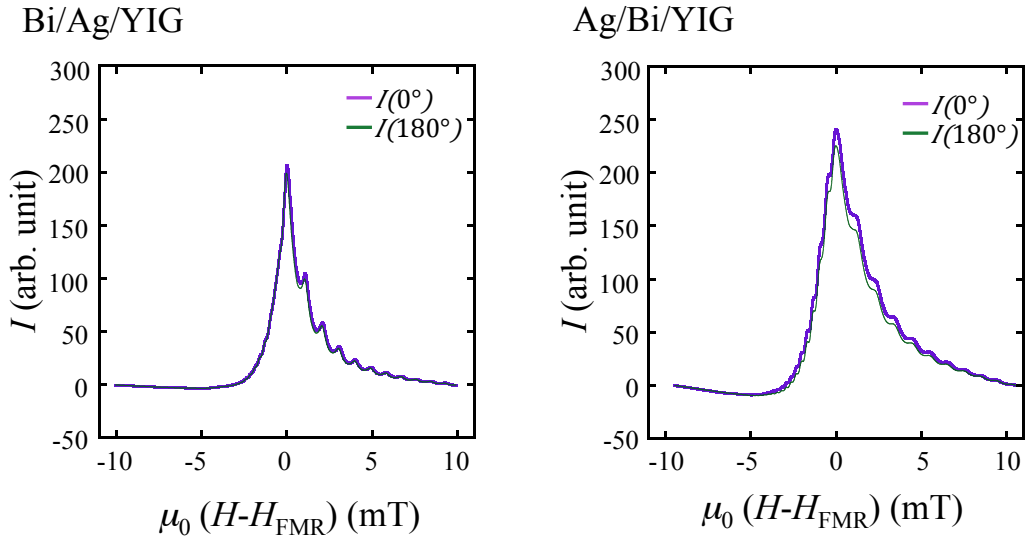

Fig. S1 FMR spectra from the Bi/Ag/YIG and Ag/Bi/YIG.

### B. Electromotive forces from Bi/Ag/Py and Ag/Bi/Py films.

To provide a material for considering a contribution of the sample quality and crystallography to the IREE, we carried out the following experiment: We prepared Bi/Ag/NiFe(Py) and Ag/Bi/Py samples on quartz substrates, and measured magnitudes of generated electric current from them. As mentioned in the main text, Py generates self-induced electromotive forces and this contribution to the generated charge current was considered for the estimation of the IREE length in them. The results are shown in Fig. S2(a). The generated electric currents of the symmetric component in the signals, which is attributed to the spin-to-charge conversion, were measured from all samples, and they are -0.59, -43 and -14 nA from Bi/Ag/Py, Ag/Bi/Py and only Py samples, respectively. When we subtract the contribution from the Py, there is the sign reversal of the electric current in the Bi/Ag/Py and the Ag/Bi/Py samples, which can suggest the existence of the IREE in the Bi/Ag interface. The IREE length is ranged from -0.095 to -0.13 nm for the $\kappa$ in the [0,1] range in our model taking into account the ISHE, the spin current reflection at the top surface of the samples and spin current transparency. One possible reason is a sample quality difference as described in the main text. In fact, if we neglect all of contributions except for the IREE, the IREE length was $+0.12 \times 10^{-5}$ nm, 5 orders of magnitude smaller than that in ref. 13 of the main text. Furthermore, the X-ray diffraction patterns of the Bi/Ag/Py and the Ag/Bi/Py on quartz substrates are largely different from those on YIG substrates (Fig. S2(b)). Especially the Bi peak was difficult to see from the Ag/Bi/Py, whereas a very small peak can be seen from the Bi/Ag/Py. The other possibility is a different magnitude of generated electric current of Py in the Py/quartz and the other samples (the Bi/Ag/Py and the Ag/Bi/Py on quartz). As reported in ref. 17, the electromotive forces, i.e., the magnitude of the generated electric current in Py were different when we changed substrates ($SiO_2$, YIG and diamond), which is attributed to different spin density gradient in the Py under the FMR due to difference of the interfacial and/or surface spin scattering rate. In our case, when Bi or Ag was grown on the Py, the spin density gradient in the Py under the FMR can be different from that in the only Py sample, and the magnitude of the electric current in the Py beneath the Bi or the Ag may be different, which may impede precise estimation of the IREE length. From these result, an introduction of a conductive ferromagnet, like Fe, makes analyses much more complex and may impede a straightforward understanding of underlying physics. We stress that our experiments using YIG allows avoiding such additional contributions.

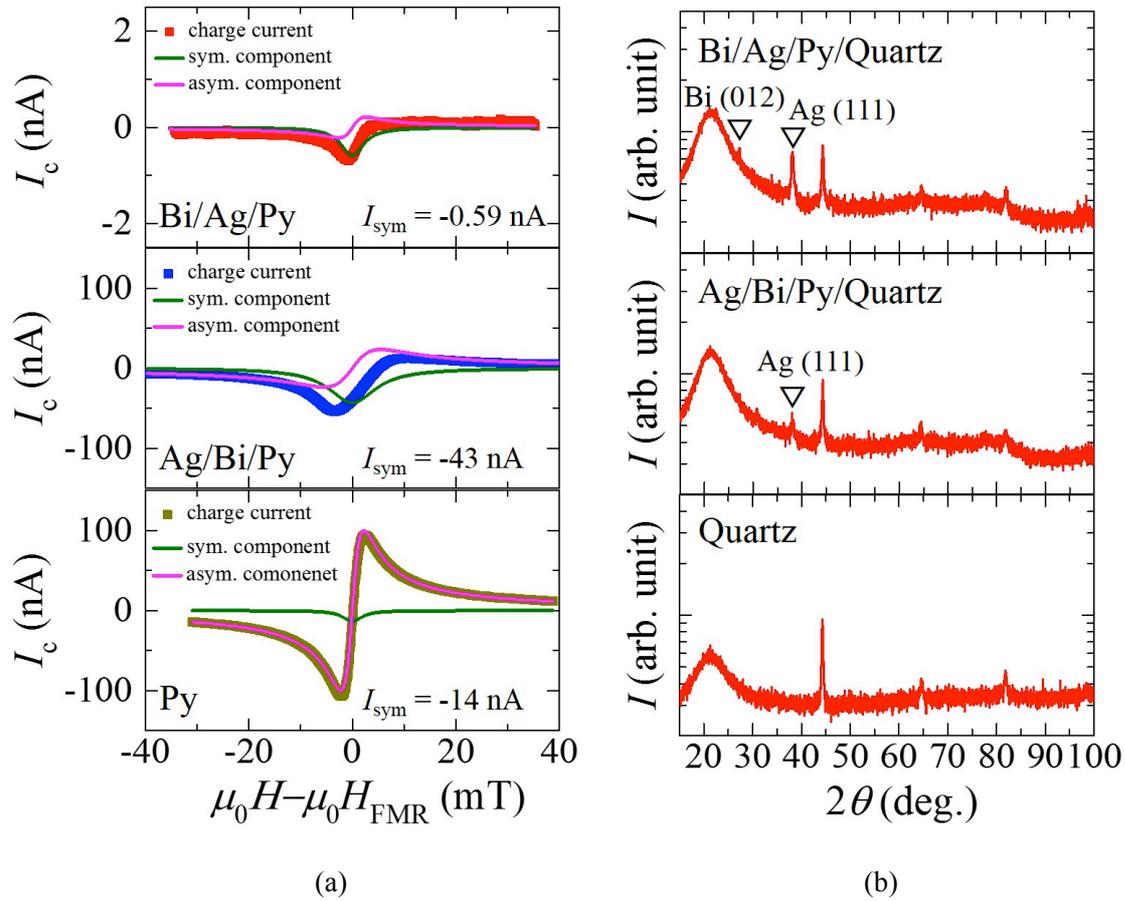

(a) (b)

Fig. S2 (a) The generated electric current from Bi/Ag/Py, Ag/Bi/Py and Py on quartz substrates. The signals were deconvoluted into a symmetric component (due to spin-to-charge conversion) and an asymmetric component (due to the anomalous Hall effect of Py), and the magnitude of the symmetric component of the electric current is shown in the figure. (b) X-ray diffraction patterns of Bi/Ag/Py, Ag/Bi/Py and Py on quartz substrates. The thickness of the Bi and the Ag was set to be the same that on YIG.